\documentclass[twocolumn,showpacs,preprintnumbers,amsmath,amssymb]{revtex4}
\usepackage{epsfig}
\newcommand{\bm}[1]{ \mbox{\boldmath $#1$}  }

\begin{document}

\title{Necessary conditions for accurate computations of three-body 
partial decay widths}

\author{E. Garrido$\:^1$, A.S. Jensen$\:^2$, D.V. Fedorov$\:^2$}
\affiliation{$^1$ Instituto de Estructura de la Materia, CSIC,
Serrano 123, E-28006 Madrid, Spain}
\affiliation{$^2$ Department of Physics and Astronomy, Aarhus University, 
DK-8000 Aarhus C, Denmark} 

\date{\today}

\begin{abstract}
The partial width for decay of a resonance into three fragments is
largely determined at distances where the energy is smaller than the
effective potential producing the corresponding wave function.  At
short distances the many-body properties are accounted for by
preformation or spectroscopic factors.  We use the adiabatic expansion
method combined with the WKB approximation to obtain the indispensable
cluster model wave functions at intermediate and larger distances.  We
test the concept by deriving conditions for the minimal basis
expressed in terms of partial waves and radial nodes.  We compare
results for different effective interactions and methods. Agreement is
found with experimental values for a sufficiently large basis.  We
illustrate the ideas with realistic examples from $\alpha$-emission of
$^{12}$C and two-proton emission of $^{17}$Ne.  Basis requirements for
accurate momentum distributions are briefly discussed.
\end{abstract}


\maketitle

\section{Introduction}

The spectrum of a given many-body quantum system provides a set of
characteristic observables: the energies and the widths.  All quantum
states, except perhaps the ground state, decay if sufficient time is
available.  For bound states, where the total energy is less than all
thresholds for division into subsystems, only electromagnetic and
particle transforming decays are possible. The first of these decays
maintains the identity of the constituent particles in contrast to the
latter exemplified by beta-decay where neutrons are transformed into
protons or vice versa. However, if the energy is sufficiently high,
the system can also decay into subsystems while maintaining the
identity of the constituent particles. In this work, we shall
concentrate on decays where the energy allows such fragmentation of
the initial system.

In general we then have a many-body continuum problem. For nuclei the
simplest final state consists of two fragments, e.g. two fission
fragments, or a daughter nucleus plus a nucleon or an $\alpha$-particle
\cite{gam28}. If both initial and final states are completely
specified, energy and momentum conservation determine the relative
kinetic energy between the two outgoing particles. The decay rate (or
the width found by multiplying the rate by $\hbar$) is obtained from
preformation, or spectroscopic factors, combined with the probability
for tunneling through the barrier in the relative potential created by
the two-body interaction.  This barrier separates the short-distance
initial many-body state from the large-distance final two-body state
\cite{kra88}.  The many-body problem is reduced to a two-body problem 
where only the two particles found after the decay appear.

Larger widths may be found in more elaborate models exploiting
different, perhaps virtual, configurations resulting in a coupled
channels problem \cite{alk03}. Such relatively simple two-body decays
have been studied from the beginning of the history of quantum
mechanics, exemplified by $\alpha$-emission \cite{gam28}.  Similar
processes vary from statistical emission of nucleons above the nucleon
separation energy for ordinary nuclei \cite{bon95}, to (almost)
instant decay outside the neutron dripline and to proton
and $\alpha$-emission outside the proton dripline \cite{tho04,bla08}.

In this work we shall consider decay processes where three fragments
are found in the final state. This is the simplest, yet not
understood, extension of the concept of two-body decay \cite{fed03}.
Furthermore, to limit the number of possible final states we assume
that the three-body threshold is lower than any other threshold.
Energy and momentum conservation still provide constraints, but the
internal distribution of the total momentum and energy between the
three fragments is not decided by these conservation laws. These
momentum and energy distributions are observables carrying detailed
information about the initial state and the process. Reliable
computations require accurate determination of the large-distance
properties of the cluster wave functions \cite{alv07,rod07}.  On the
other hand, the decay width is an average quantity which is very
sensitive to properties of the potential barrier, but in analogy to
two-body decay, it is determined by the effective barrier at small and
intermediate distances.

Nowadays, different methods to compute the partial decay width into
three specified fragments are already available
\cite{gar04,rod07,gri00}. However, the conditions for their reliability and
suitability are not well established, and each method is most often
rather tested on the individual systems under investigation. The
discussions comparing the different methods are deceivingly mixing
effects of choices of (i) degrees of freedom, (ii) interactions, (iii)
theoretical method, and (iv) numerical convergence. Untangling these
effects is badly needed to formulate necessary conditions for accurate
computations. Benchmark computations for precisely specified systems
and interactions would be valuable as test criteria for reliability of
the methods.

The purpose of the present paper is to provide verifiable simple but
revealing test examples.  To do this it is necessary to separate and
assess the impact of each of the effects (i)-(iv).  We first explain
why the basic ingredient for partial three-body decay widths
necessarily must be a three-body cluster model. We shall formulate
necessary conditions for accurate three-body computations, and
document by numerical applications on realistic three-body decaying
systems.  It is crucial that extensions to include more complicated
effects are built on methods which are established as accurate.

The basic concepts for three-body decay widths are in section 2
described and tested against measured results.  In section 3 we give
analytical estimates of the quantities characterizing the crucial
potentials and the basis size needed for accurate computations. In
section 4 we test the estimates with the clean example of the Hoyle
resonance in $^{12}$C.  We then discuss in section 5 how the computed
widths depend on the basic interactions and the available methods.  We
also briefly discuss the more severe accuracy requirements for
momentum distributions determined at larger distances.  Finally
section 6 contains summary and conclusions.

\section{Framework}

Different definitions exist of the decay width, e.g. in terms of cross
sections \cite{tay72}, or phase shifts and $S$-matrix poles
\cite{fri91}, or eigenvalues of the Hamiltonian $H$ \cite{lan65}.  We
shall use the width of a given resonance defined as minus two times
the imaginary part of the generalized eigenvalue of the Hamiltonian.
This is equivalent to a pole in the $S$-matrix at the complex momentum
corresponding precisely to that eigenvalue.  This definition is only
complete after specification of the degrees of freedom contained in
$H$.  The full many-body Hamiltonian would give the total width, while
confined to three particles in the final state the result is an
approximation to the corresponding partial decay width.

\subsection{Basic concept}

The important issue in three-body decays is that the many-body degrees
of freedom must be (re)organized into intrinsic and relative cluster
coordinates.  This division is usually not meaningful at small
distances when all particles are close and within the nuclear volume.
In contrast, this is the only meaningful division at large distances
where three free particles are present.  The quantum mechanical wave
function must reflect this transition and the three-body structure at
large distance must unfold into a many-body structure at small
distance.  To get a correct partial three-body width, effects from the
small distances must be incorporated, e.g. by a preformation factor
and the assumption of an artificial attractive pocket designed to
provide both the correct energy and the resonance small-distance
boundary condition.  This treatment is precisely as for
$\alpha$-emission in the classical Gamow theory
\cite{gam28,kra88,fed03}.

In practice the many-body problem is therefore transformed into a
three-body problem for all distances.  The potential pocket in
three-body coordinates has to be added by hand, unless the chosen
two-body interactions already are sufficient.  It is crucial to have
the correct three-body resonance energy as evident from the
exponential energy dependence of the width determined from tunneling
probability through any barrier.  Thus we have to insist on a
practical method to adjust the energy to the correct value, e.g. by
use of a short-range three-body interaction \cite{fed96}.  This
separates the model dependence of short-distance many-body structure
from effects of the three-body cluster model at intermediate
distances.  In Gamow theory this is achieved by adding an attractive
$\alpha$-daughter interaction, e.g. a square-well or Woods-Saxon
potential with a radius about the size of the nucleus.

All methods to compute partial three-body decay widths must at some
point address this separation of distances, degrees of freedom and
related effective interactions.  However, the procedures to reach the
reduction into the three-body structure are rather different.
Microscopic derivations start from nucleon-nucleon interactions and
integrate away the unwanted degrees of freedom while simultaneously
leaving corresponding effective interactions
\cite{kam07,des01,ara03,kan07,nef04}.  The mean-field model is also
used for two-proton decay in the formulation named the Gamow
shell-model where the nucleon degrees of freedom are present and the
interactions are phenomenological adjusted within the mean-field
\cite{mic02,bet02,oko03}.

If the three-body cluster model is assumed from the beginning for all
distances, the corresponding two-body interactions are usually
obtained from phenomenological adjustments to two-body data.  No
matter how the effective two-body interactions are found, they are
much more important in three-body than in two-body decays. For example
for $\alpha$-emission, the distance between $\alpha$-particle and
daughter quickly leaves only non-vanishing contributions from Coulomb
and centrifugal forces \cite{alv07,rod07}.  For three-body decay, where two
particles stay close while the third particle moves away, the
short-range interactions contribute much more to the properties of the
decisive barrier \cite{gar04}.

The crux of the matter is then to deduce the partial decay width from
the three-body cluster model for a given resonance energy.  Here
several methods have been employed.  The practical and experiment
oriented method is to use elastic scattering cross sections as
function of energy \cite{tay72,fri91}. A peak is then related to a
resonance and its width is the resonance width.  However, this is not
practical for collisions of more than two particles, and furthermore
uncertainties arise from corrections due to phase space distortion,
broad peaks, overlapping resonances, background contributions, etc.
We prefer the more mathematical definitions of a complex pole in the
$S$-matrix \cite{fri91}, or the energy derivative of the scattering
phase shift while crossing through $\pi/2$ \cite{tay72,fri91}, or
equivalently minus twice the imaginary part of the eigenvalue of the
hamiltonian \cite{lan65}, or equivalently related to the solution of
complex energy with an outgoing flux in all channels.

The numerical method to compute the complex resonance energy has to
allow for complex energy solutions and for example insist on only an
outgoing flux in all channels \cite{ara96}.  Equivalently all coordinates
can be rotated a given angle into the complex plane which turn the
outgoing flux solution into an exponentially falling solution at large
distances precisely as for bound states \cite{ho83,cso97}.  Another method
exploits analytic continuity of the interactions by varying with
respect to for example the strength parameter \cite{tan99,tan99b}.  These
methods are essentially all equivalent and produce identical results
for identical interactions and the same Hilbert space. 

A semiclassical perturbative method has been abundantly employed
recently \cite{gri00,gri03,gri07}.  It assumes that the width is small
and related to the outgoing flux for the solution to the
Schr\"{o}dinger equation obtained by confinement to a box of finite
extension.  This method is presumably also equivalent to the other
methods for narrow resonances when the perturbation assumption is
valid.

\subsection{Testing the concept}

The bare problem can be envisaged as a potential and a corresponding
wave function with resonance character conveniently described by a
complex energy and only an outgoing flux.  Real ($E_R$) and imaginary
($E_I$) parts of the energy are closely linked through the potential,
e.g. adding a small positive potential at small distance would
increase $E_R$, leave the potential barrier unchanged, and increase
$|E_I|$ exponentially.  We shall test the idea with the simplest
computation of the width, $\Gamma = - 2 E_I$, i.e. by use of the
WKB-approximation.

However, to do this, a potential defined as function of a generalized radial 
coordinate is required.  The coordinates for three
particles then should be combined into one overall important
coordinate $\rho$ which on its own should be able to describe the
process.  We want to maintain the intuitive understanding that small
$\rho$ means small physical distances between all three particle at
the same time.  Vice versa,  large $\rho$ should be able to describe the
structure after fragmentation into three particles.  To give a precise
and physically meaningful definition of $\rho$ in terms of the
particle coordinates we assume a quadratic dependence on interparticle
distances.  Then the hyperradius $\rho$ is unique apart from the
choice of mass weighting, where we use $mM\rho^2 = \sum_{i<j} m_im_j
(\vec r_i -\vec r_j)^2$, with $m_i$ and $\vec r_i$ as mass and
coordinate of particle $i$, and $M=m_1+m_2+m_3$ \cite{mac68,nie01}.
In our calculations we take the normalization mass
$m$ equal to the nucleon mass. 

We shall use $\rho$ as the generalized coordinate. In the spirit of
the generator coordinate method, where the energy is calculated as
function of one parameter, we solve the Schr\"{o}dinger or Faddeev
equations for each value of $\rho$ between zero and $\rho_{max}$. The
hyperangular degree of freedom is automatically quantized as the
solution to the corresponding Schr\"{o}dinger equation, which for
each $\rho$ takes the form:  
\begin{equation}
\hat{\Lambda}^2 \Phi_n(\rho,\Omega)+\frac{2 m \rho^2}{\hbar^2} V(x)
\Phi_n(\rho,\Omega)  = \lambda_n(\rho) \Phi_n(\rho,\Omega),
\end{equation}
where $\hat{\Lambda}$ is the hyperangular operator \cite{nie01}, $V$ is 
the sum of the three two-body potentials, and $\Omega$ represents the 
five hyperangles.

If we now use the complete set of solutions for each $\rho$ as basis 
($\Psi={1\over\rho^{5/2}} \sum_n f_n(\rho) \Phi_n(\rho,\Omega)$)
we arrive at the hyperspherical adiabatic expansion method, which reduces 
the problem to the following set of coupled differential equations in $\rho$: 
\begin{eqnarray}
\left[ -\frac{d^2}{d\rho^2} +  \frac{2m}{\hbar^2} (V_{3b}(\rho) - E)
+ \frac{1}{\rho^2}
\left( \lambda_n(\rho)+\frac{15}{4} \right) \right] f_n(\rho) \nonumber &&\\
+ \sum_{n'} \left( -2 P_{n n'} \frac{d}{d\rho} - Q_{n n'} \right)f_{n'}(\rho)
= 0, \;\;\; & & \label{eq0}
\end{eqnarray}
where $E$ is the three-body energy, $V_{3b}$ is a three-body potential used 
for fine-tuning, and the functions $P_{n n'}$ and $Q_{n n'}$ are given for 
instance in \cite{nie01}.

These equations contain the effective adiabatic potentials, which take the
form:
\begin{equation}
V_{\mbox{eff}}(\rho)=\frac{\hbar^2}{2m}
    \frac{\lambda_n(\rho)+15/4}{\rho^2}+V_{3b}(\rho).
\end{equation}
The number of differential equations (or of adiabatic potentials) is equal 
to the size of the basis for each $\rho$.  This basis is unique in the description of
multifragmentation, because it is the only representation that maps
fragmentation theory onto a set of coupled channel differential
equations from two-body reaction theory
\cite{mac02,mac68}. 

To fix the concept we shall first only use the lowest of these
adiabatic potentials which depend on angular momentum and parity of
the three-body system.  The three-body configurations change in a
non-trivial manner from small to large hyperradii. The total energy,
apart from the kinetic energy related to variation of $\rho$, is
minimized for each $\rho$.  However, $\rho$ does not uniquely
determine even the geometric configuration.  The same $\rho$ is
related to continuously differing combinations of distances between
the particles, e.g.  a small distance between two particles and a
large distance to the third particle, or equal distance between all
particles, etc.  Thus, it is not a priori obvious why this should be a
good choice of coordinates for an efficient description of these
decays.  Other configurations might be important but this would be
reflected in a finite population of the complete set of higher-lying
adiabatic potentials.

\begin{figure}
\begin{center}
\vspace*{0.3cm}
\epsfig{file=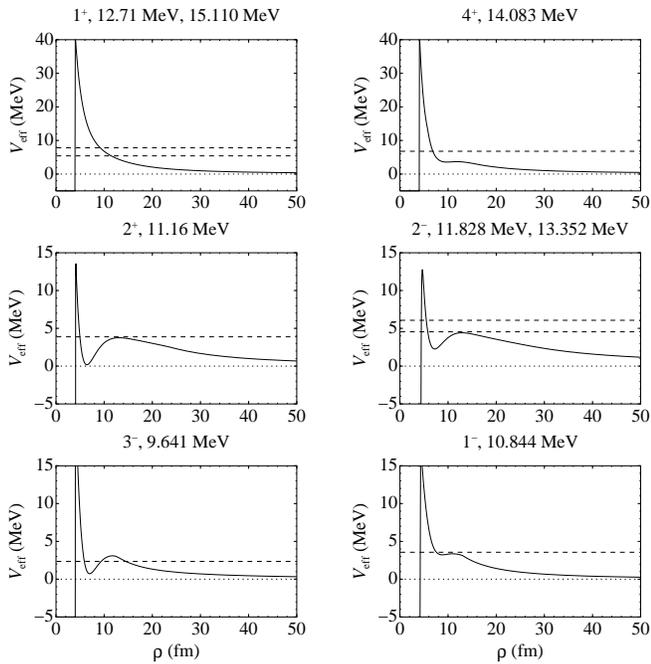,scale=0.45,angle=0}
\end{center}
\vspace*{-0.5cm}
\caption{The computed lowest adiabatic potential as function of hyperradius
for a number of resonances of $^{12}$C ($\alpha$+$\alpha$+$\alpha$).
The horizontal lines mark the resonance energies measured above the
three-body threshold.  The corresponding excitation energies are
given above each of the panels \cite{ajz90}.  }
\label{fig1}
\end{figure}

In Fig.\ref{fig1} we show the lowest adiabatic potentials for various angular 
momenta and parities $J^{\pi}$ in $^{12}$C ($\alpha$+$\alpha$+$\alpha$).
The behavior of the three body systems is unpredictable from $J^{\pi}$
alone.  A minimum at small distances indicates a substantial amount of
cluster structure and vice versa, no minimum indicates dominance of
many-body non-cluster structure often referred to as shell model
structure. In all cases, after decay, the three alpha particles must emerge
outside the barrier.  The physical meaning of this description is
different from the intuitive perception of a mean-field shell model
where a monotonous dependence on excitation energy and angular
momentum would be expected. Here the three-body structure and the
properties of the two-body interactions are crucial and capable of
changing the expected ordering.

The partial decay width must sensitively depend on the energy and the
properties of the barrier which must be crossed.  The complex energy
solutions of the hyperradial equation give the widths, which we here
estimate by the WKB tunneling probabilities through the
one-dimensional potential barriers.  We fix the real part of the
energies equal to the measured values whenever they are available.
This can be achieved by using an appropriate three-body short-range
attractive potential ($V_{3b}(\rho)$ in Eq.(\ref{eq0})). A simple form of 
this is employed in
Fig.\ref{fig1}, where we take a square well potential of radius
$4$~fm and a depth adjusted to give the experimental resonance
energies.

\begin{figure}
\begin{center}
\vspace*{0.3cm}
\epsfig{file=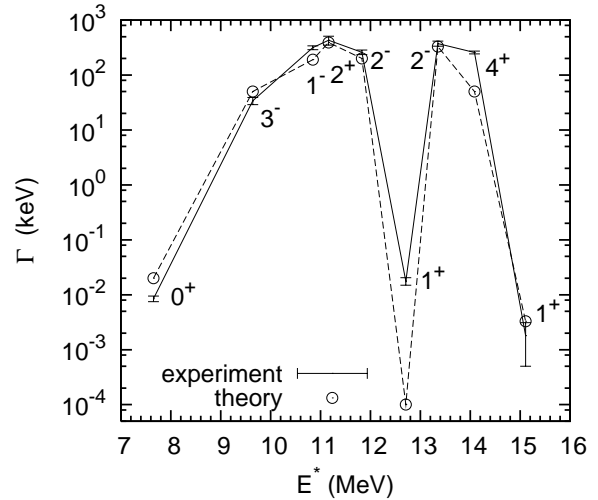,scale=0.66,angle=0}
\end{center}
\vspace*{-0.5cm}
\caption{The computed widths compared to the measured values for
different resonances of $^{12}$C with excitation energy $E^*$
\cite{ajz90}. The WKB approximation is used for tunneling through the
lowest adiabatic potential where the inner turning point is fixed to a
hyperradius of $\rho = 4$~fm. }
\label{fig2}
\end{figure}

The WKB estimates are obtained from the action integral between the
classical turning points determined by the real part of the energy 
($E$). The transmission coefficient is then given by:
\begin{equation}
T=\exp{\left\{-2 \int_{\rho_i}^{\rho_t} \left[
\frac{2 m}{\hbar^2}(V_{\mbox{eff}}(\rho)-E)
                           \right]^{1/2} d\rho \right\} } \; ,
\end{equation}
where $\rho_i$ and $\rho_t$
are the inner and outer classical turning points defining the distance
through the barrier. Once the transmission coefficient is computed, the 
decay constant can be obtained as $\Gamma /\hbar$=$f T$, where $f$ is the
knocking rate.

The results are shown in Fig.\ref{fig2} where only a factor of 2-5
remains to match the measured values which vary by about 5 orders of
magnitude with only a variation of 7~MeV in excitation energy $E^*$. The
smallest widths occur for the smallest angular momenta but the largest
widths do not occur for the largest angular momenta.
The only exception is the lowest $1^+$ state where the computed	width is 
too small by about a factor of $100$.  The reason is that the
large-distance tail of the corresponding potential sensitively
contributes to the width for this state, see Fig.\ref{fig1}.  The
basis size should then in this case have been larger than used to
obtain this estimate.  

In more accurate computations, as discussed in details in this report,
the inner turning point would also be larger implying larger computed
widths. This would in turn indicate reduction due to preformation, or
spectroscopic, factors, arising from a short-distance structure
deviating from that of three $\alpha$-particles.  In any case, we can
conclude that the use of the hyperradius as the generating coordinate
is valid to account for the main dependencies of the resonance widths
on the excitation energy and angular momentum and parity of the
system.

\section{Crucial ingredients}

The exponential dependence of the widths on the properties of the
confining barrier immediately emphasizes that the most crucial
ingredient is the potential barrier which must be accurately
determined between classical turning points.  Any inaccuracies would
be exponentially enhanced in the width computation.  Thus, it is
essential first to know the coordinate region where accuracy is
indispensable which means that the turning points should be found.  Second,
it is crucial to control the numerical technique responsible for the
computation of the potential barriers in this coordinate range.

\subsection{Turning point estimates}

The inner turning point is at a distance where the short-range
interaction is most important.  It is sensitive to the details of the
potential and the energy and quantum numbers of the resonance.
However, the short distance allows accurate computation without too
many difficulties, provided that the cluster division is assumed and
the many-body problem reduced to that of three particles.

Then the outer classical turning point is an estimate of the largest
distance needed in accurate computations.  If we assume that only
repulsive Coulomb interactions remain at these distances we can use
hyperradial coordinates and find the turning point $\rho_t$ with the
formalism developed in \cite{gar05}.  

When a direct decay of the three-body system is assumed, all relative
two-body distances scale proportionally. We can then minimize the
action integral and arrive at the general expression \cite{gar05}
\begin{equation}\label{e50} 
 \rho_t = \frac{e^2}{E\sqrt{mM}} \left(
   \sum_{i<k} (z_i z_k)^{2/3} (m_i m_k)^{1/3} \right)^{3/2} 
\end{equation}
where $e z_i$ is the charge of particle $i$ and $E$ is the energy
of the three-body resonance which at the turning point equals the 
potential energy.  

The partial waves necessary to describe this distance can be estimated
from the fact that at the turning point the value of the potentials
can not be higher than the three-body energy $E$. Therefore, the
centrifugal barrier must be smaller than $E$ at the turning point,
i.e.
\begin{equation} \label{e60}
 \frac{\hbar^2}{2m} (\frac{\ell_x(\ell_x+1)}{x^2} + 
 \frac{\ell_y(\ell_y+1)}{y^2} ) < E \;,
\end{equation}
where $x$ and $y$ are the usual Jacobi coordinates, and $\ell_x$ and
$\ell_y$ are relative angular momenta related respectively to the
distance between two particles and their center of mass and the third
particle.  Assuming that both the total angular momentum and the
intrinsic spins are small, we have $\ell_x \approx
\ell_y$, and for sufficiently large values of  $\ell_x$ and  $\ell_y$
we can take $\ell_x(\ell_x+1)\approx \ell_x^2$ and
$\ell_y(\ell_y+1)\approx \ell_y^2$. Furthermore $x$ is proportional to
the distance between two of the particles, say $1$ and $2$, and for a
direct decay all distances between pairs of particles are similar.
This leads to
\begin{equation} \label{e70}
 x \approx \rho_t \sqrt{\frac{M \mu_{12}}{ \sum_{i<k} m_i m_k}} \;\;\; , \;\;\;
 y \approx  \sqrt{\rho_t^2 -x^2}
\end{equation}
with $\mu_{12} = \frac{m_1 m_2}{m_1+m_2}$, from which one gets
\begin{equation} \label{e80}
 \ell_x \approx \ell_y < \frac{\rho_t}{\hbar \sum_{i<k} m_i m_k}  
 \sqrt{2 E m M m_1 m_2 m_3}  \:.
\end{equation}
Two limits of Eqs.(\ref{e50}) and (\ref{e80}) are useful in
practice. When all particles have the same masses $m_0$ and charges
$z_0$ we find
\begin{eqnarray} \label{e90}
 \rho_t \approx  \frac{3 z_0^2 e^2}{E} \sqrt{\frac{m_0}{m}}  \;\; ,\;\;
 \ell_x \approx \ell_y <  z_0^2 e^2  \sqrt{\frac{6m_0}{\hbar^2 E}}  \; . 
\end{eqnarray}
When particle $3$ has mass $m_3$ and charge $z_3$ much larger than for
the other two particles, i.e. $m_1 \approx m_2 \ll m_3$ and $z_1
\approx z_2 \ll z_3$, we obtain
\begin{eqnarray} \label{e100}
 \rho_t \approx \frac{z_1 z_3 e^2}{E} \sqrt{\frac{8 m_1}{m}} \;\; ,\;\; \ell_x
 \approx \ell_y < 2 z_1 z_3 e^2  \sqrt{\frac{m_1}{\hbar^2 E}} \; .
\end{eqnarray}
Notice here that the estimates of lengths in hyperspherical
coordinates always are combined with the normalization mass $m$, e.g.
only $\rho_t \sqrt{m}$ is expressed in terms of physical parameters
like particle energies, masses and charges.  In contrast dimensionless
quantities like the angular momenta in Eqs.(\ref{e90}) and
(\ref{e100}) do not involve the arbitrary mass $m$.

Instead of following a path where all distances scale proportionally
it could be advantageous to tunnel through the barrier by exploiting
the two-body attraction between two particles while the third particle
moves away.  This is a sequential decay where the geometry reduces the
Coulomb interactions and supplies additional energy from the
short-range interaction.  In total the barrier could be substantially
smaller both in height and width.  If the intermediate two-body
configuration carries an energy $E_{12}$ we find
\begin{eqnarray} \label{e110}
 \rho_t \approx \frac{z_3 (z_1+z_2) e^2}{(E -E_{12})} 
 \sqrt{\frac{m_3(m_1+m_2)}{mM}}  \;  \\  \label{e115}
  \ell_y < z_3 (z_1+z_2) e^2 \sqrt{\frac{2 m_3(m_1+m_2)}
 {M \hbar^2 (E -E_{12})}} \; .
\end{eqnarray}
Where we have used that for sequential decay large values of $\rho$
imply $\rho \approx y$, and the relative distances $r_{31}$ and
$r_{32}$ are similar to $r_{12,3}=y \sqrt{m/\mu_{12,3}} \approx \rho
\sqrt{m/\mu_{12,3}}$.  The angular momentum estimate is an upper limit
because the two-body attraction can typically only be exploited for
one or very few given small values of $\ell_x$ which by angular
momentum conservation also forces $\ell_y$ to have a small value.

In the two limits of the same masses ($m_0$) and charges ($z_0$), and
one mass and charge ($m_3$ and $z_3$) much larger than the other two
($m_1\approx m_2$ and $z_1 \approx z_2$), we find that
\begin{eqnarray} \label{e120}
 \rho_t \approx \frac{ 2 z_0^2  e^2}{(E -E_{12})}  
 \sqrt{\frac{ 2 m_0 }{3 m }} \;\;,
 \; \; \ell_y <  \sqrt{\frac{16 m_0 (z_0^2 e^2)^2} 
 {3 \hbar^2 (E -E_{12})}} \;,\\ \label{e125}
 \rho_t \approx  \frac{ 2 z_3 z_1 e^2}{(E -E_{12})} 
 \sqrt{\frac{ 2 m_1}{ m}}  \;\;, \;\;  
 \ell_y <  \sqrt{\frac{16 m_1 (z_3 z_1 e^2)^2} { \hbar^2 (E -E_{12})}} \; .
\end{eqnarray}

The balance between the smaller Coulomb energy and the additional
two-body energy $E_{12}$ determines whether the tunneling process is
direct (all pairs of particles are outside their short-range
attraction at the turning point and all distances scale
proportionally) or sequential (the attraction from one or more pairs
of particles is used as vehicle to speed up the process by tunneling
through a smaller barrier).  Different paths can contribute to the
same decay process.  The dominating path would typically correspond to
the smallest outer turning point $\rho_t$.

\subsection{Basis requirements}

The width can only be accurately computed if the potential and the
related wave functions are accurate up to at least the true outer
turning point.  For this it is important that the Hilbert space is
sufficient to allow the system to choose the optimum path through an
accurately determined barrier.  This in turn can only be achieved when
the numerical procedure allows a precise coherent description of all
two and three-body intermediate configurations.  

For coherently contributing two-body substructures this is only
ensured by Faddeev, or Faddeev-like, decompositions.  One set of
Jacobi coordinates is obviously simpler than including all three
Faddeev components but it also has a tremendous disadvantage for
systems where more than one two-body subsystem simultaneously have
bound or nearly bound states, and especially when relatively large
distances are important.  The Efimov effect is for example completely
excluded from the description in a basis of one Jacobi set of coordinates.

In many computations basis expansions are exploited.  With the
hyperradius as coordinate the natural basis is hyperharmonics for each
Jacobi set in the basis.  To describe structures varying over
distances comparable to the range of the two-body interaction,
$b_{12}$, the basis must on average contain a few points within that
distance. For a given $\rho$ the two-body distance is
$r_{12}=\sqrt{m/\mu_{12}} x$ ($x \equiv \rho \sin\alpha$), which means
that a proper description of the internal two-body structures for
large $\rho$'s requires basis terms with nodes within the small region
$0\leq \alpha \leq \sqrt{\mu_{12}/m}\; b_{12}/\rho$. The number of
nodes between 0 an $\pi/2$ of a given hyperspherical harmonic is given
by the principal quantum number $n$ of the Jacobi polynomial contained
in it. This means that a typical separation between nodes is
$\pi/(2n)$.  Therefore, to have at least one node below $
\sqrt{\mu_{12}/m}\; b_{12}/\rho$ one needs
\begin{equation}
n > \frac{\pi}{2}\frac{\rho}{b_{12}}\sqrt{\frac{m}{\mu_{12}}},
\end{equation}
which implies that the maximum hypermomentum quantum number $K_{max}$
($=2n+\ell_x+\ell_y$) should at least exceed a lower limit, i.e.
\begin{eqnarray} \label{e130}
 K_{max} >  \frac{\pi \rho}{b_{12}} \sqrt{\frac{m}{\mu_{12}}} +\ell_x+\ell_y
 \; ,
\end{eqnarray}
where we assumed given hyperradius and partial angular momenta.  Thus
the number of basis functions, or $K_{max}$, must increase
proportional to $\rho$.  The proportionality factor providing the unit
of $\rho$ for this estimate is inversely proportional to the range of
the two-body interaction which initially is responsible for the
structures we try to describe. The minimum basis size can only be found in 
practice by numerical calculations.  

When the turning point, $\rho=\rho_t$, is needed we can insert the
expression in Eq.(\ref{e50}) in Eq.(\ref{e130}).  Then only the
physical parameters remain in the estimate of $K_{max}$.  Furthermore,
to relate $\rho_t$ to the physical size of the system the definition
of the hyperradius can be used, which leads to $m \rho_t^2
\approx r_{av}^2 \sum_{i<k} m_i m_k /M$ where $r_{av}^2$ is an appropriate 
average of the distance, $(\vec r_i -\vec r_j)^2$, between pairs of
particles.

\section{A test case: The Hoyle state in $^{12}$C.}

We consider here the Hoyle state, i.e. the first $0^+$ resonance in
$^{12}$C described in terms of three $\alpha$-particles, which is the
only possible particle decay mode.  The intrinsic $\alpha$-particle
spins are zero and the wave function must be symmetric with respect to
all interchanges of pairs of particles.  This limits the number of
necessary partial waves and the $K_{max}$-dependence of the effective
potential and wave function can conveniently be investigated.

The well-studied converged result is known to be a narrow resonance
with a width of only 8.5 eV at an energy of 0.38 MeV.  A detailed
calculation using the hyperspherical adiabatic expansion method can be
found in \cite{rod07} where the structure is shown to arise
essentially from only one of the adiabatic potentials.

Following the estimates in Eq.(\ref{e90}) where a direct decay is assumed,
we obtain an estimated value for the outer turning point of 
$\rho_t \approx 90$~fm, while the maximum values for the two-body relative
orbital angular momenta are $\ell_x\approx\ell_y\approx 7$. However,
the proportional scaling of all distances is inefficient because the two-body 
attraction is not exploited while the full Coulomb repulsion is encountered.  
By using Eq.(\ref{e120}) instead we find $\rho_t \approx 67$~fm with $E_{12} =
0.1$~MeV as the energy of the $^{8}$Be subsystem, and $\ell_y\approx 8$.  
The coherence of all three subsystems reduces the turning point 
and keeps the number of components.  Then the barrier is reduced 
and this escape mechanism is preferred. For the second case ($\rho_t \approx 
67$~fm and $\ell_y\approx 8$), and  following (\ref{e130}), we get that $K_{max}$ 
has to be about 80 in order to ensure a proper treatment of the $^{8}$Be structure
up to the outer turning point.

\begin{figure}
\begin{center}
\vspace*{-1.2cm}
\epsfig{file=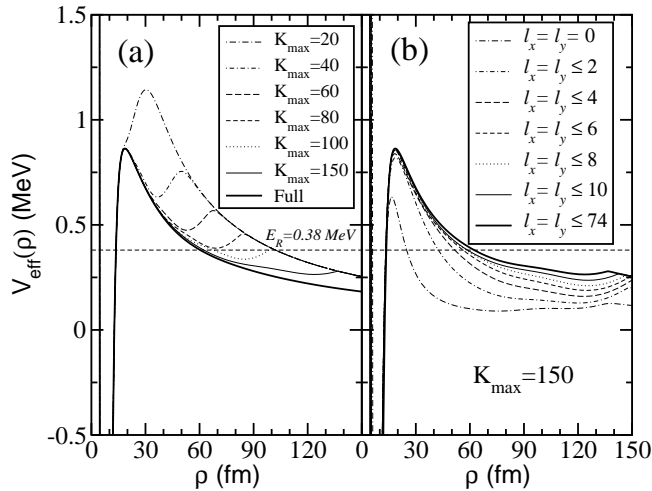,scale=0.37,angle=-90}
\end{center}
\vspace*{-1.0cm}
\caption{The dominating effective adiabatic potential for the lowest 
0$^+$ resonance in $^{12}$C ($\alpha$+$\alpha$+$\alpha$) for (a) different
values of $K_ {max}$ and (b) different values of partial waves in the expansion 
in terms of HH. The dashed straight line indicates the energy of the
resonance. In (a) all the the possible values of $\ell_x$ and $\ell_y$ 
consistent with $K_{max}$ have been included. The curve called ``full"  
includes $\ell_x = \ell_y$ up to 12 only and $K_{max}$=150, but with $K_{max}$ 
increased up to 500 for some of the components. In (b) $K_{max}$ is taken equal to 150. }
\label{fig3}
\end{figure}

In the left part of Fig.\ref{fig3}, we show the lowest adiabatic effective 
potential for the $0^+$ states in $^{12}$C for different values
of $K_{max}$. The dashed straight line indicates the energy of the
resonance. In this calculation all the possible values of $\ell_x$ and of 
$\ell_y$ consistent with a given $K_{max}$ have been included. The curve 
quoted as full has been computed with $K_{max}$=150 and values of $\ell_x$ 
and $\ell_y$ up to 12 only, but increasing $K_{max}$ up to 500 for the 
most contributing components.
As seen in the figure, the computed potential barrier (and the outer 
turning point) changes dramatically with $K_{max}$, and, in agreement with 
the estimate of Eq.(\ref{e130}), a $K_{max}$ value of at least 80 is needed 
in order to obtained a converged potential up to the outer turning point, which is
found to be 60 fm (also in agreement with the estimated value).

In the right part of the figure we show the same effective potential for 
a fixed value of $K_{max}$ (we have taken 150), but where the values of 
$\ell_x$ and $\ell_y$ are progressively increased. It is clear from the figure that
too small values of $\ell_x$ and $\ell_y$ produce a too small potential 
barrier. As estimated above, values of $\ell_x$ and $\ell_y$ of at least 
8 are needed to match the thick curve, which is the one plotted in the 
right part with $K_{max}=150$ and with values of $\ell_x$ and $\ell_y$ 
of up to 74.

It is then clear from Fig.\ref{fig3} that simultaneous use of appropriate
$K_{max}$ and maximum values of $\ell_x$ and $\ell_y$ are sufficient
to reproduce accurately the potential barrier up to the outer turning 
point. This is the decisive region of the potential determining the width
of a given resonance. It is important to note that while a too small value
of $K_{max}$ overestimates the potential barrier, too small values of
$\ell_x$ and $\ell_y$ underestimate it. Therefore both effects
tend to compensate each other, in such a way that a poor calculation using 
too low $K_{max}$ and too few $\ell_x$ and $\ell_y$ could however luckily 
give a computed width not too far from the correct one.

\begin{table}
\caption{WKB estimates for the width of the lowest 0$^+$ resonance in $^{12}$C
for the seven effective potentials shown in Fig.\ref{fig3}a (left part of the
table) and for the seven potentials shown in Fig.\ref{fig3}b (right part of the
table). The column labeled as $\Gamma_{pert}$ refers to the widths obtained with the perturbative
method described in \cite{gri00}. All the widths are given in MeV. The experimental value is 
$8.5 \cdot 10^{-6}$ MeV.}
\label{tab1}
\begin{tabular}{|cccc|ccc|}
  \hline
$K_{max}$& &    $\Gamma_{WKB}$    &  $\Gamma_{pert}$     &$\ell_x$,$\ell_y$ & & $\Gamma_{WKB}$  \\ \hline
  20     & & $3.6 \cdot 10^{-11}$ &  $1.7\cdot 10^{-12}$ &   $=0$           & & $3.1 \cdot 10^{-3}$  \\
  40     & & $3.9 \cdot 10^{-10}$ &  $1.4\cdot 10^{-11}$ &   $\leq 2$       & & $6.8 \cdot 10^{-5}$  \\
  60     & & $2.7 \cdot 10^{-9}$  &  $9.1\cdot 10^{-9}$  &   $\leq 4$       & & $1.5 \cdot 10^{-5}$  \\
  80     & & $2.7 \cdot 10^{-8}$  &  $3.9\cdot 10^{-8}$  &   $\leq 6$       & &$7.9 \cdot 10^{-6}$   \\
  100    & & $3.0 \cdot 10^{-6}$  &  $2.7\cdot 10^{-7}$  &   $\leq 8$       & &$6.3 \cdot 10^{-6}$   \\
  150    & & $4.2 \cdot 10^{-6}$  &  $7.4\cdot 10^{-6}$  &   $\leq 10$      & &$5.4 \cdot 10^{-6}$   \\
  Full   & & $5.5 \cdot 10^{-6}$  &  $5.2\cdot 10^{-6}$  &                  & &                      \\  \hline
\end{tabular}
\end{table}

The dependence of the width on the basis size is illuminating.  In
table \ref{tab1} we give the computed WKB widths for the different
effective potentials shown in Fig.\ref{fig3}a (left part of the table)
and Fig.\ref{fig3}b (right part of the table). The conclusion is
striking, an insufficient basis (small $K_{max}$ or small maximum
values of $\ell_x$ and $\ell_y$) easily leads to widths overestimated
or underestimated by $3-5$ orders of magnitudes. However the converged
result agrees reasonably well with the experimental value of $8.5$
eV. One has to keep in mind that only the uncertainty arising from the
different possible choices of the knocking rate when computing the WKB
width can easily produce variations in the width of up to a factor of
2 or 3. In our calculations we have taken the knocking rate equal to
the energy of the resonance divided by the Plank constant
($E=h\nu$). Furthermore, in this estimate we have neglected the higher
lying adiabatic potentials which typically contribute about 10-15\% of
the wave function. In any case, the width is exponentially depending
on the barrier, and we have therefore demonstrated how catastrophic it
is to use an insufficient basis.

A sufficiently large basis for three-body quantities is already
important for structure computations of weakly bound halo systems
\cite{jen04}.  Excitations into the continuum of two-neutron halos, 
where the Coulomb interaction is absent, give more sensitivity and
the opportunity to compare computations with measured strength
functions. The inadequacy of a small basis was exhibited in
\cite{fed97} for the $1^-$ excitation of the $^6$He ground state.
This was further emphasized in the fairly successful prediction of the
similar $1^-$ strength function for $^{11}$Li in \cite{gar02} which
later on was accurately measured in \cite{nak06}.

\section{Interactions and methods}

For $\alpha$-emission it is well-known that the long-range
interactions (as the Coulomb potential) and the centrifugal forces, 
by far have the largest influence on the decay rates. This is less obvious for
three-body decays since the potential barriers depend on the two-body interactions. 
On the other hand the decay
rate should be independent of the method adopted for the computations.
This does of course assume that the methods are mathematically and/or
physically equivalent which not always is easy to confirm.

\subsection{Interaction dependence}

An asymmetric and numerically more difficult system is found in
$^{17}$Ne ($^{15}$O+$p$+$p$), which provides an example of an
exceedingly long-lived two-proton decaying state.  This nucleus has a
resonance with spin and parity 3/2$^-$ with an energy of 0.34 MeV
above threshold. The ground state of $^{16}$F ($^{15}$O+$p$) is a
resonance at 0.53 MeV above threshold. Decay of the 3/2$^-$ resonance
in $^{17}$Ne via a $^{16}$F resonance is then not allowed due to
energy conservation.  Also, core mass and charge are substantially
larger than the values of the valence particles, meaning that the
estimates of Eqs.(\ref{e100}) are applicable. From them we get $\rho_t
\approx 96$~fm and $\ell_x < 6$.  From Eq.(\ref{e130}) we then
estimate that $K_{max} \approx \rho_t/$(1~fm), i.e. to reach an accurate
turning point when the two-body channels are exploited we can expect
$K_{max}$-values of about 100.

The numerical difficulties are related both to less symmetry due to
the presence of non-identical particles and the non-zero intrinsic
spins of $1/2$ of all three particles.  The number of partial waves
are then immediately rather large due to the many allowed spin
couplings. The most important interactions for the resonance structure
are the $s$, $p$ and $d$-waves of the proton-$^{15}$O system, each
producing two low-lying resonances in $^{16}$F.

\begin{figure}
\begin{center}
\vspace*{-1.2cm}
\epsfig{file=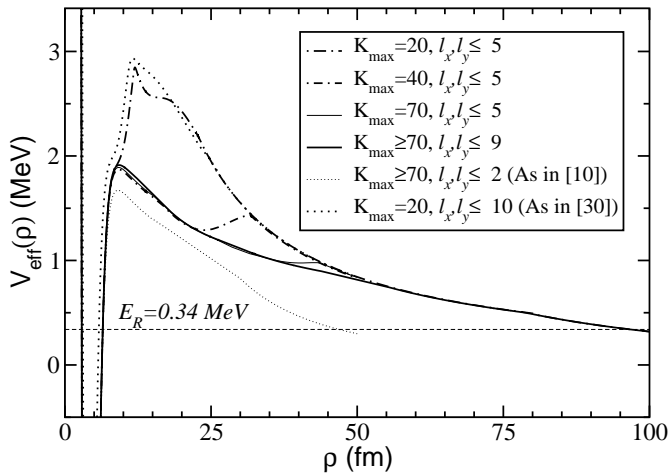,scale=0.36,angle=-90}
\end{center}
\vspace*{-1.0cm}
\caption{Lowest adiabatic effective potential for the 3/2$^-$ resonance 
in $^{17}$Ne. Except for the last curve (thick-dot), the $p$-$^{15}$O
interaction is the one described in \cite{gar04}. The different curves
correspond to calculations with different values of $K_{max}$ and
relative two-body angular momenta. The thin-dotted curve corresponds
to the calculation presented in \cite{gar04}. In the last
(thick-dotted) curve, the $p$-$^{15}$O potential and $K_{max}$ value
of ref.\cite{gri03} have been used.  The dashed straight line indicates
the resonance energy of 0.34~MeV.}
\label{fig4}
\end{figure}

The lowest adiabatic effective potential for the 3/2$^-$ resonance is
shown in Fig.\ref{fig4}. Except for the last case in the figure
legend (thick-dotted curve) all the calculations have been performed
using the interactions in \cite{gar04}. We notice the same pattern as
in Fig.\ref{fig3}, i.e. increasing $K_{max}$ up to values of about
$70$ is required for convergence of the effective potential up to the
outer turning point. This turning point appears at about 90 fm, in
agreement with the estimation from Eq.(\ref{e100}). Otherwise the
barrier is overestimated, giving rise to too small resonance
widths. Also, the number of partial waves needed to obtain a proper
convergence is relatively small ($\ell_x$ and $\ell_y$ values not
bigger than 5 is enough, also in agreement with the estimation from
Eq.(\ref{e100})). If only $s$, $p$, and $d$-waves are used,
the barrier (as illustrated by the thin-dotted curve) is too small, and
consequently the computed width is too big. This barrier is the one used
in ref.\cite{gar04} where a resonance width of $3.6\cdot 10^{-12}$ MeV
was obtained. The true width is therefore expected to be a few orders
of magnitude smaller than this number.

The thick dotted curve in Fig.\ref{fig4} has been obtained with
$K_{max}=20$ and including all the partial waves consistent with this
value, in total corresponding to the same Hilbert space as in
\cite{gri03}.  This curve is also obtained with the same interactions
as in \cite{gri03} where a width of $4.1\cdot 10^{-16}$ MeV is quoted.
This width is consistent with the $5.4 \cdot 10^{-16}$ MeV obtained
with the WKB approximation for this potential.  However, as seen in the
figure the barrier is substantially overestimated compared to the
results with a much larger basis.  The implication is that the width
should be bigger than this number.

In Fig.\ref{fig4}, the thick solid line corresponds to a calculation
with $\ell_x, \ell_y \leq$ 9, and $K_{max}$=70 for all the partial
waves, except for those of large amplitude where $K_{max}$ values
between 130 and 160 are used.  Then the potential barrier shows the
correct convergence properties and the WKB width for this potential
and the resonance energy of $0.34$~MeV is found to be $1.7 \cdot
10^{-14}$ MeV.  This value is, as expected, in between the two given
in \cite{gar04} and \cite{gri03}, and consistent with the corrected
width of $(5-8) \cdot 10^{-15}$ MeV obtained in \cite{gri07} for a
larger $K_{max}$ value.  As seen in the figure, an increase of
$K_{max}$ up to 40 improves significantly the calculated effective
potential, although the barrier is still a bit overestimated. It it
important to keep in mind the limitations inherent to the WKB
approximation. Only one adiabatic potential is included, and
additional inaccuracy arises from the definition used for the knocking
rate and the fact that a preformation factor of unity is used.

The potentials shown in Fig.\ref{fig4} are to a large extent
determined by the Coulomb and centrifugal potentials. This is
understandable, since many of the crucial properties are determined at
distances larger than the ranges of the short-range interactions. To
investigate the dependence on these two-body interactions we can
compare the results arising from the two potentials obtained with
$K_{max}$=20. The first of them (long-dashed-dot curve) has been
obtained with the proton-$^{15}$O interaction given in \cite{gar04},
while the second one (thick-dot curve) uses the one in \cite{gri03}.
These two interactions are very different, especially in their
parametrization of the spin-dependence.  In \cite{gar04} the spin-spin
and spin-orbit operators are $\bm{s}_c \cdot \bm{j}_p$ and
$\bm{\ell}\cdot \bm{s}_p$, where $\bm{j}_p=\bm{\ell}+\bm{s}_p$,
$\bm{s}_c$ and $\bm{s}_p$ are the spins of the core and the proton,
and $\bm{\ell}$ is their relative orbital angular momentum.  In
\cite{gri03} the more symmetric form of $\bm{s}_c \cdot \bm{s}_p$ 
and $\bm{\ell}\cdot(\bm{s}_c+\bm{s}_p)$ is used.

The second set of spin-dependent operators in \cite{gri03} does not
preserve the usual mean field quantum numbers corresponding to the
core, and is as such inconsistent with the description of the protons
in $^{15}$O.  This symmetry breaking is especially problematic in
cases where one spin-orbit partner is occupied by core nucleons while
the other is available for the valence nucleons.  Then the $d_{5/2}$
and $d_{3/2}$ resonances in $^{17}$Ne are mixed in the two-body
description of the $1^-$ and $2^-$ resonances of $^{16}$F.  In the
three-body problem of $^{17}$Ne the two valence protons are then
forced to partly occupy the same orbits.  The Pauli principle between
core and valence protons is violated and unwanted properties may
appear, see \cite{gar03} for details.  To compensate for these effects
a fully phenomenological $J^{\pi}$ dependent three-body potentials are
used but their effects on the widths could be rather unpredictable.
An example of unwanted properties is seen in \cite{gri03}, where the
potentials obtained with symmetry breaking spin operators are adjusted
with three-body potentials to reproduce the known two-body spectrum in
$^{16}$F.  However, this simultaneously results in an additional $2^-$
resonance in $^{16}$F at 2.8~MeV with a width of 0.26~MeV.  This
resonance and the corresponding bound state in $^{16}$N at $-0.38$~MeV
are not mentioned in the description of the interaction \cite{gri03},
and furthermore they are not known experimentally.

In Fig.\ref{fig4} it is shown how such exceedingly different two-body
interactions lead to effective potentials in almost perfect agreement,
provided that the same basis is used. Only a small difference is found
around the inner turning point which is more sensitive to properties
of the short-range interactions.  This strongly indicates that the
specific values of the energies of the two-body resonances are the
only decisive quantities for the intermediate distances corresponding to
the lowest adiabatic effective potential.  Small-distance properties
essential for spectroscopic factors as well as dynamic evolution of
the resonance structures with distance due to couplings between
potentials could in contrast be sensitive to the design of the
interactions.  These properties are determined inside or outside the
barrier, respectively.

The conclusion from these examples is that the width is surprisingly
insensitive to the two-body interactions as long as they provide the
proper attraction for the lowest adiabatic potentials.  This
conclusion has two crucial assumptions,  i.e. first the three-body
resonance energy is adjusted to the correct energy by use of a
short-range three-body potential. Second, the effective barrier is
accurately computed for example by use of a sufficiently large basis
where it is important to allow the higher partial waves in the Hilbert
space although the corresponding two-body interactions do not have to
be precise.  The explanation is simply that Coulomb potential and
centrifugal barriers dominate at the intermediate distances where the
confining barrier is located and the width in turn is determined. In
cases where this turns out to be incorrect the width may depend much
more on the specific choice of the two-body interactions.  The most
tempting guess of such a situation is a large width corresponding to a
narrow barrier and an outer turning point at a relatively small
distance.

\subsection{Method dependence}

The methods to calculate resonances defined as poles of the $S$-matrix
must all make use of analytic continuation into the complex plane.
This inevitably involves an approximation to the physical quantity of
interest, e.g. the imaginary part or the resonance width is not found
in a physical process and therefore not directly an observable.  The
connection has to be established through theoretical derivations and
model dependent interpretations.  The continuation is only possible
when an analytical form is available as for example the potentials in
the Schr\"{o}dinger equation.  If they are given as numerical tables
obtained by fits to measured cross sections at discrete points, the
numbers must be connected by an analytical expression.  This is the
same result as if analytical parametrized potentials from the
beginning are adjusted to reproduce experimental values.  Either way
the potentials can be continued into the complex plane.  

The further away from the real axis, the larger is the uncertainty
arising both from the model dependent interpretations in terms of
observables and from the somewhat arbitrary choice of the initial
analytical form. Thus large computed widths are for these reasons
intrinsically more uncertain than the small ones.  These methods are
equivalent if the same potentials are employed.  The choice of method
is then only a matter of numerical, and perhaps mathematical,
convenience.  One possibility is the use of complex scaled coordinates,
i.e. all lengths in the Schr\"{o}dinger (or Faddeev) equation are
multiplied by $\exp(i\theta)$, where $\theta$ is a given angle which
has to be larger than half of the angle of rotation from the real
energy axis to the direction of the resonance defined in the complex
energy plane.  Then the resonance wave function in the rotated space
is an eigenfunction with bound state boundary conditions for the
corresponding complex energy \cite{ho83}.

Rotating the resonance solution back to the real coordinates results
in a wave function with only outgoing flux and the same complex
energy.  This boundary condition and a complex energy without complex
coordinates are then fully equivalent.  A third method is to continue
the interactions analytically by varying a strength parameter
\cite{tan99,tan99b}.  The results are then obtained as function of this parameter
and in the end extrapolated back to the correct physical value. An
example is the computation of the broad $0^+$-resonance in $^{12}$C
\cite{kur07}.  However, the large width is already an indication of an 
inherent uncertainty.  For all these methods the computed widths are
model dependent approximations most accurate for relatively small
widths.

An almost identical method is the Gamow shell model which so far only
is applicable for nucleons and a core \cite{oko03,mas07}.  It is the
complex rotated mean-field shell model with wave functions expanded on
an ordinary single-particle basis. The computed eigenvalues are
investigated as functions of the number of basis states and their
spatial extension.  A few of the eigenvalues converge to specific
complex energies corresponding to resonance positions and widths,
while the majority only reflect an attempt to discretize the
continuum. Only uncorrelated motion can be described.

It is also possible to remain on the real axis, both energy and
coordinates, but the resonance energy and width must be defined in a
different way.  In a scattering formulation the elastic cross section
varies through a peak over a small range of energy.  Apart from
quantitatively important and qualitatively unimportant corrections,
the peak position and its width are the resonance energy and the
width.  A large width then means that the cross section is smeared out
and resembling the background which obviously makes both energy and
width determination more uncertain.

Another definition of the width is through the time dependent, or
stationary wave packet of an evolving wave function.  An initial
condition then has to be assumed, e.g. a source term in the
Schr\"{o}dinger equation adding probability at small distances.  The
decay rate by which this probability disappears is then directly
interpreted as the resonance width.  When more resonances contribute
this is much more complicated and turns into a coupled channels
problem.  Analyses of experimental data use this formulation expressed
as transition probabilities through potential barriers corresponding
to two consecutive coherent two-body decays \cite{dig05}.

The simplest version of a stationary incident wave packet attempting
to tunnel through a barrier is estimated by the semiclassical
WKB-tunneling probability.  This formulation is still possible for a
multi-channel problem, simply by choosing a one-dimensional path
through the many-dimensional space.  By definition this is a
semiclassical method where the tunneling probability should be small
and the potentials should be smooth.

\subsection{Perturbation treatment}

Another possibility is the perturbation treatment used in
\cite{gri00}.  The width is obtained in three steps. First
the bound state problem is solved in a box of hyperradius $\rho_b$
less than the outer turning point $\rho_t$. Second, this wave function
is used as a source term of arbitrary strength to find the resonance
wave function, and third the decay rate (or the width) is found by computing
the outgoing flux at an arbitrary large $\rho_{max}$ distance. At this
stage it is then necessary to impose the proper boundary condition to the
computed wave function, which for a system of three charged particles
is not known explicitly. A detailed discussion about possible different 
approximations to implement the Coulomb boundary condition in the hyperspherical 
harmonic expansion method can be found in \cite{gri01}. In our adiabatic
approach it is enough to impose to the radial solution to go as 
${\cal G}_\xi(\eta,\rho)+i{\cal F}_\xi(\eta,\rho)$,
where ${\cal F}$ and ${\cal G}$ are the Coulomb functions, and the index
$\xi$ and the Sommerfeld parameter $\eta$ are obtained numerically
from the adiabatic potential (see \cite{gar07} for details).  

The assumption is that the perturbatively computed wave function describes
the true resonance, and the related outgoing flux at the corresponding
$\rho_{max}$ radius gives the decay rate.  The first of the steps does
not require knowledge of the details of the outer part of the
barrier. However, calculation of the outgoing flux at a given
$\rho_{max}$ value beyond the barrier obviously requires barrier
knowledge at least up to $\rho_{max}$.

Therefore, for this procedure to work it is necessary either to
compute accurately the potential barrier up to $\rho_{max}$ or to make
some assumption about the neglected outer parts of the barrier, which
still are inside the turning point.  This is unavoidable since an
infinitely thick barrier smoothly continued from the box radius
$\rho_b$ inside the turning point would give vanishing decay rate
(width equal to zero). The opposite assumption of vanishing barrier
outside the box would result in a finite decay rate. This means that
if the barrier is overestimated, as for too small $K_{max}$, the
outgoing flux at $\rho_{max}$ will be reduced producing a too small
width.  It is then meaningless to expect an accurate computation
of the tunneling probability through a barrier without accurate
knowledge of that barrier.

In the same way, one should be very careful when computing observables
obtained from the behaviour of the wave function at very large values
of the hyperradius (as for instance 1000 fm, as quoted in
\cite{gri07,gri03}).  Direct computation of these large-distance 
properties is in obvious conflict with the poor knowledge of the outer
barrier. If typically one needs a large value of $K_{max}$ to get
accurate calculations up to the outer turning point, this $K_{max}$
value should be clearly bigger if accurate results are required for
distances of a few times $\rho_t$.

The procedure applied in \cite{alv07,rod07} circumvent this problem of
direct calculation at unreachable large distances.  First it is
tempting to employ momentum space instead of coordinate space but this
would at best only change the problem into uncertainties at large
momenta or equivalently the small distances decisive for the resonance
structure would be uncertain.  If both small and large distances are
needed accurately it is equally convenient to work in coordinate space
where the small distances naturally are most accurately computed.  To
get accurate large-distance asymptotic properties, the desired
observables should be computed with acceptable accuracy for a given
large basis at the largest possible distance.  The computed observable
must then remain unchanged when the distance is further increased
provided the basis is correspondingly increased.  Passing this test is
equivalent to sufficient accuracy at both small and asymptotically
large distances.

Comparison of different methods is quantified in table~\ref{tab1}
where the third column, $\Gamma_{pert}$, gives the widths of the Hoyle
resonance in $^{12}$C computed using the perturbative method described
in \cite{gri00} with inclusion of only the first adiabatic potential,
precisely as for the WKB results, $\Gamma_{WKB}$, in the second column.  We
note that the perturbative results are consistent with the ones
obtained in the WKB approach, although they tend to be about one order
of magnitude smaller, especially when $K_{max}$ is far from the one
required for convergence. Again, calculations made with a too small
value of $K_{max}$ give rise to too small widths, while the
calculation made with a sufficiently large $K_{max}$ and a
sufficiently large number of partial waves provides a result in good
agreement with the WKB estimate and the experimental value. The values
obtained with the perturbative method depend weakly on the
$\rho_{max}$ value and the value of $\Gamma$ used to obtain the wave
function with outgoing boundary condition. In any case, the computed
results do not change significantly.

Similar agreement is obtained for the 3/2$^-$ state in $^{17}$Ne. In
this case the values of $\Gamma_{pert}$ oscillate between the
$4.9\cdot 10^{-17}$ MeV obtained with $K_{max}$=20 (long-dashed-dot
curve in Fig.\ref{fig4}), and the $3.0\cdot 10^{-15}$ MeV obtained
with $K_{max} \leq 70$ (thick solid line in the figure). For these two
cases the WKB estimates were $5.4\cdot 10^{-16}$ MeV and $1.7\cdot
10^{-14}$ MeV, respectively.

Therefore, the different methods give rise to similar widths, at least
within the method uncertainties. Once the three-body energy of the resonance
is given correctly (with the help for instance of an effective 
three-body force), the two methods are equally sensitive to the deficiencies
of the potential barriers arising from the use of a poor basis.
Use of a too small $K_{max}$ value or too few partial waves produces
incorrect barriers, and therefore uncontrolled widths.

\begin{figure}
\begin{center}
\vspace*{-1.2cm}
\epsfig{file=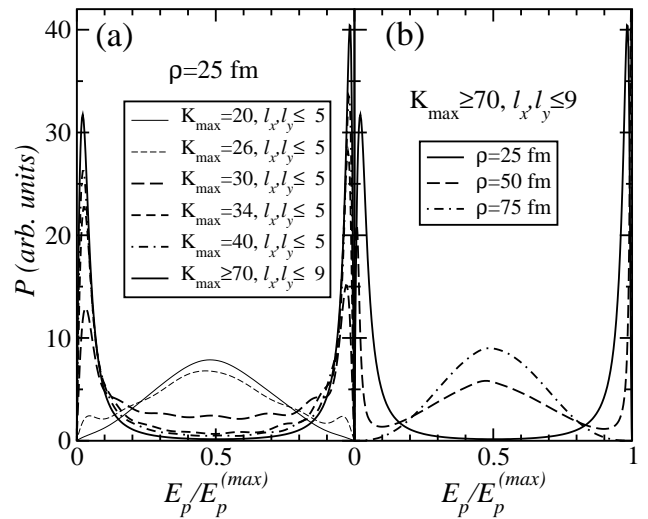,scale=0.38,angle=-90}
\end{center}
\vspace*{-0.8cm}
\caption{Proton energy distributions for the 3/2$^-$ resonance in $^{17}$Ne
where the maximum allowed emission energy is used as unit.  The left
part (a) is for $\rho$=25 fm for different values of $K_{max}$.  The right
part (b) is for a very large basis for increasing hyperradii.  Only the
lowest adiabatic effective potential is used.}
\label{fig5}
\end{figure}

\subsection{Fragment energy distribution }

After three-body decay of a given resonance, one of the most
investigated observables is the energy distributions of the fragments
after the decay.  As shown in \cite{gar06}, these quantities are
mainly determined from the properties of the coordinate space wave
functions at large distances. This happens because the hyperspherical
harmonics transform into themselves after Fourier transformation into 
momentum space. The kinetic energy distribution of the fragments at a given 
$\rho$ is therefore, except for a phase-space factor, obtained as the absolute
square of the total wave function in coordinate space for that $\rho$,
but where the five hyperangles are interpreted as in momentum space.

To obtain the energy distributions after decay one obviously needs to
consider a value of $\rho$ clearly beyond the potential barrier, which
means values of the hyperradius of several times the outer turning
point.  Following the discussion in the previous section, it is clear
then that an accurate computation of such distributions requires a
basis able to describe properly the three-body state at such large
distance. In other words, reliable calculations require sufficiently
large values of $K_{max}$, which will be larger than the ones
reproducing the potential barrier up to the outer turning point.

For a system like $^{17}$Ne, where the outer turning point for the
3/2$^-$ resonance is about 90 fm, the basis needed to compute the
energy distributions after decay becomes soon very big, and therefore
the numerical calculation very heavy.  However, to illustrate the
importance of a sufficient size of the basis, we first show in left
part of Fig.\ref{fig5} the energy distribution of the proton when
different $K_{max}$ values are used and when $\rho$ is fixed at 25
fm. For such $\rho$ value, as seen in Fig.\ref{fig4}, accurate
calculations are possible. Although this value of $\rho$ is clearly
below the outer turning point, it illustrates the dependence of the
energy distributions on the basis size. The only thing to have in mind
is that the plotted distributions correspond to an intermediate stage,
before the decay of the resonance. As a consequence, the plotted
distributions do not necessarily resemble the experimental ones, which
should be computed at a much larger value of $\rho$. The interplay
between the two-body interactions can clearly change the energy
distributions when moving from small to large values of $\rho$.

In any case, from Fig.\ref{fig4}, we know that at $\rho$=25 fm a
$K_{max}$ value of at least 40 is needed to have a converged potential
at this distance. This fact is also reflected in left part of
Fig.\ref{fig5}, where we see that the energy distribution obtained
when $K_{max}$=40 matches pretty well the result obtained with a much
bigger basis (thick solid line). This energy distribution corresponds
to a situation in which one of the protons takes most of the energy,
while the other one stays close to the core. However, when the basis
size decreases the proton energy distribution begins to fill the
intermediate region, such that when $K_{max}$=20 (thin solid line)
only one wide peak at about 0.5 is seen. The same strong dependence on
$K_{max}$ as the one observed in the figure can be found for larger
values of $\rho$, in particular for the values needed when computing
the energy distributions after the decay.

To see the dependence on the hyperradius we show in the right part of
Fig.\ref{fig5} how the proton energy distribution changes with the
largest basis used in the left part.  Now increasing $\rho$ shifts the
proton energy distribution from two narrow peaks at low and high
energy to one broad peak around half the maximum energy.  These
distributions have not yet converged as function of $\rho$.  A larger
value is needed with a correspondingly larger basis but the trend is
clear.  The protons are emitted roughly with equal energy.  It is very
illuminating to notice that the correctly converged result obtained
for a large hyperradius and a large basis resembles an inaccurate
result from a small basis and a much too small hyperradius.

\section{Summary and conclusions }

The partial width for a resonance decaying into three fragments must
necessarily deal with the corresponding three-body problem.  As for
$\alpha$-emission we believe the crucial ingredients are provided by
effective potentials at intermediate and large distances measured
relative to the radius of the decaying nucleus.  The many-body effects
essentially only enter at small distances as preformation or
spectroscopic factors and in microscopic derivations of the effective
two-body interactions.

We concentrate in this paper on the three-body problem.  We assume
that the small distance boundary conditions necessary for resonance
computations is simulated by the correct three-body energy and an
attractive pocket within the confining barrier.  We furthermore assume
it is possible to specify the effective two and three-body
interactions.  With these assumptions we have isolated the crucial
three-body problem from the underlying many-body degrees of freedom.
Accurate resonance structures, partial decay widths, and fragment
momentum distributions must first of all be computed for this well
defined few-body problem.  Inclusion of many-body effects may be
important but this cannot avoid the three-body problem.  The more
elaborate schemes only add to the complexity involved in obtaining
three-body input parameters.

We first formulate the basic concept of effective three-body
potentials.  As a test we then compare experimental widths for a
number of $^{12}$C resonances with results from the simplest WKB
application.  As for $\alpha$-emission the largest variation is
reproduced only leaving effects from preformation factors or
equivalently many-body effects amounting to less than one to two
orders of magnitude.  The reproduced systematics is not monotonous
with excitation energy or angular momentum.

The essence of the problem is now narrowed down to accurate
computations of the effective three-body potentials. The classical
turning points for the dominating potential therefore specify
the region of interest in coordinate space. Any uncertainty in this
region is enhanced exponentially in the computed widths.  We give
analytical estimates of both turning points and the number of contributing
partial waves.  We also estimate analytically, and demonstrate
numerically, which basis size is necessary for accurate computations of
the potentials in this coordinate region.  Almost all published
results employ insufficient basis sets.

We investigate the dependence of the computed widths on the effective
two-body interactions.  As for $\alpha$-emission the total energy is
crucial and the two-body interactions should only provide roughly the
same resonance energies independent of contributing spin and orbital
angular momentum structure.  The decisive properties are supplied by
Coulomb and centrifugal forces.  We discuss the different methods
applied to width computations and conclude that they in most cases are
equivalent.

Finally we illustrate how it is more difficult to get accurate
momentum distributions of the fragments after three-body decay.  These
observables are sensitive to properties of the coordinate space wave
functions at distances outside the turning point. Therefore an even
larger basis is required.  Unfortunately, a two-fold inaccurate
computation with a too small basis and a too small hyperradius
by a strange coincidence resembles the correctly converged result.

In conclusion, the partial three-body widths for decay of many-body
resonances are first of all determined by three-body properties.
Many-body effects are less important.  The effective three-body
potentials or equivalently the corresponding three-body wave functions
are decisive and must be accurately computed.


\begin{thebibliography}{99}


\bibitem{gam28} G. Gamow, Z. Phys. {\bf 51}, 204 (1928); {\bf 52}, 510 (1928).

\bibitem{kra88} K.S. Krane, {\it Introductory Nuclear Physics}, 
John Wiley \& Sons (1988) Page 254. 

\bibitem{alk03} J. Al-Khalili and F. Nunes, J. Phys. G {\bf 29}, R89 (2003).

\bibitem{bon95} J. P. Bondorf, A. S. Botvina, A. S. Iljinov, I. N. Mishustin, and K. Sneppen,
Phys. Rep. {\bf 257}, 133 (1995).

\bibitem{tho04} M. Thoennessen, Rep. Prog. Phys. {\bf 67}, 1187  (2004).

\bibitem{bla08} B. Blank and M.J.G. Borge, Progr.Part.Nucl.Phys., 
{\bf 60}, 403 (2008).

\bibitem{fed03} D.V. Fedorov, H.O.U. Fynbo, E.Garrido, and A.S. Jensen,
Few-Body Systems {\bf 34}, 33 (2004).

\bibitem{alv07} R. \'{A}lvarez-Rodr\'{\i}guez, A.S. Jensen,  D.V. Fedorov,
H.O.U. Fynbo, and E. Garrido, Phys.Rev.Lett. {\bf 99}, 072503 (2007).

\bibitem{rod07} R. \'{A}lvarez-Rodr\'{\i}guez, E. Garrido, A.S. Jensen, 
D.V. Fedorov, and H.O.U. Fynbo, Eur. Phys. J. A {\bf 31}, 303 (2007).

\bibitem{gar04} E. Garrido, D.V. Fedorov, and A.S. Jensen, 
Nucl. Phys. A {\bf 733}, 85 (2004).

\bibitem{gri00} L.V. Grigorenko, R.C Johnson, I.G. Mukha, I.J. Thompson,
and M.V. Zhukov,  Phys. Rev. Lett.  {\bf 85}, 22 (2000).

\bibitem{tay72} J.R. Taylor Scattering Theory, Wiley and Sons, 
New York 1972, chapter 20.

\bibitem{fri91} H. Friedrich,  Theoretical Atomic Physics, 
Springer-Verlag, Berlin Heidelberg 1991, chapter 4.

\bibitem{lan65} L.D. Landau and E.M. Lifshitz, Quantum mechanics, 
Pergamon Press, 1965, chapter XVII paragraph 134.

\bibitem{fed96} D.V. Fedorov and A.S. Jensen, Phys.Lett. B {\bf 389}, 631 (1996).

\bibitem{kam07} R. Kamouni and D. Baye, Nucl.Phys. A {\bf 791}, 68 (2007).

\bibitem{ara03}  K. Arai, P. Descouvemont, D. Baye, W.N. Catford, 
Phys.Rev. C {\bf 68}, 014310 (2003). 

\bibitem{des01}  P. Descouvemont and A. Kharbach, 
Phys.Rev. C {\bf 63}, 027001 (2001).

\bibitem{kan07}  Y. Kanada-En'yo, Prog.Theor.Phys. {\bf 117}, 655 (2007). 

\bibitem{nef04}  T. Neff and H. Feldmeier, Nucl.Phys. {\bf 738}, 357 (2004).

\bibitem{oko03} J. Okolowicz, M. Ploszajczak, and I. Rotter, Phys. Rep. {\bf 374}, 271 (2003).

\bibitem{mic02} N. Michel, W. Nazarewicz, M. Ploszajczak, and K. Bennaceur, 
Phys. Rev. Lett. {\bf 89}, 042502 (2002). 

\bibitem{bet02} R. Id Betan, R. J. Liotta, N. Sandulescu, and T. Vertse, 
Phys. Rev. Lett. {\bf 89}, 042501 (2002). 

\bibitem{ara96}  K. Arai, Y. Ogawa, Y. Suzuki, K. Varga,
Phys.Rev. C {\bf 54}, 132 (1996).

\bibitem{cso97}  A. Csoto, R.G. Lovas and A.T. Kruppa, 
Phys.Rev.Lett. {\bf 70}, 1389 (1993); 
A. Csoto and G.M Hale, Phys.Rev.C {\bf 55}, 536 (1997).

\bibitem{ho83} Y.K. Ho, Phys. Rep. {\bf 99}, 1 (1983).

\bibitem{tan99} N. Tanaka, Y. Suzuki, K. Varga, and R.G. Lovas, 
Phys.Rev. C {\bf 59}, 1391 (1999).

\bibitem{tan99b} N. Tanaka, Y. Suzuki, and K. Varga, Phys.Rev. C {\bf 56}, 562 (1997).

\bibitem{gri07} L.V. Grigorenko and M.V. Zhukov, Phys. Rev. C {\bf 76}, 014008 (2007).

\bibitem{gri03} L.V. Grigorenko, I.G. Mukha, and M.V. Zhukov, Nucl. Phys. A {\bf 713},
                372 (2003).

\bibitem{mac68} J.H. Macek,  J.Phys., {\bf 1}, 831 (1968). 

\bibitem{nie01} E. Nielsen, D.V. Fedorov, A.S. Jensen and E. Garrido, 
Phys. Rep. {\bf 347}, 373 (2001).

\bibitem{mac02} J.H. Macek,  Few-Body Systems, {\bf 31}, 241 (2002).

\bibitem{ajz90} F. Ajzenberg-Selove, Nucl. Phys. A {\bf 560}, 1 (1990),
and http://www.tunl.duke.edu/nucldata/chain/12\underline{ }newv.shtml.

\bibitem{gar05} E. Garrido, D.V. Fedorov, A.S. Jensen, and H.O.U. Fynbo, 
Nucl. Phys. A {\bf 748}, 27 (2005).

\bibitem{jen04} A.S.~Jensen, K. Riisager, D.V.~Fedorov, and E. Garrido, 
Rev. Mod. Phys. {\bf 76}, 215 (2004).

\bibitem{fed97} D.V. Fedorov, A. Cobis, and A.S. Jensen,
Phys. Rev. C {\bf 59}, 554 (1999).

\bibitem{gar02} E. Garrido, D.V. Fedorov, and A.S. Jensen, 
Nucl. Phys. A {\bf 708}, 277 (2002). 

\bibitem{nak06} T. Nakamura et al., Phys. Rev. Lett. {\bf 96}, 252502 (2006).

\bibitem{gar03} E. Garrido, D.V. Fedorov, and A.S. Jensen, 
Phys. Rev. C {\bf 68}, 014002 (2003).

\bibitem{kur07}  C. Kurokawa and K. Kato,  Nucl.Phys. {\bf A 792} (2007) 87;
Phys. Rev. C {\bf 71}, 021301(R) (2005).

\bibitem{mas07} H. Masui, K. Kato, and K. Ikeda, 
Phys. Rev. C {\bf 75}, 034316 (2007).
 
\bibitem{dig05} C. Aa. Diget et al., Nucl. Phys. A {\bf 760}, 3 (2005).

\bibitem{gri01} L.V. Grigorenko, R.C. Johnson, I.G. Mukha, I.J. Thompson, 
and M.V. Zhukov, Phys. Rev. C {\bf 64}, 054002 (2001).

\bibitem{gar07} E. Garrido, D.V. Fedorov, A.S. Jensen, and H.O.U. Fynbo,
Nucl. Phys. A {\bf 790}, 96 (2007). 

\bibitem{gar06} E. Garrido, D.V. Fedorov, A.S. Jensen, and H.O.U. Fynbo,
Nucl. Phys. A {\bf 766}, 74 (2002). 



\end{thebibliography}
\end{document}